\def\spose#1{\hbox to 0pt{#1\hss}}
\def\approxlt{\mathrel{\spose{\lower 3pt\hbox{$\sim$}}
        \raise 2.0pt\hbox{$<$}}}
\def\approxgt{\mathrel{\spose{\lower 3pt\hbox{$\sim$}}
        \raise 2.0pt\hbox{$>$}}}

\def\multleft#1{\hbox to size{\vbox {\halign {\lft{##}\cr #1}}\hfill}\par}
\def\multright#1{\hbox to size{\vbox {\halign {\rt{##}\cr #1}}\hfill}\par}

\def\degmark{^\circ}
\def\boxit#1{\vbox{\hrule\hbox{\vrule\kern3pt\vbox{\kern3pt
          #1 \kern3pt}\kern3pt\vrule}\hrule}}

\def\cm{{\rm\thinspace cm}}

\def\erg{{\rm\thinspace erg}}

\def\K{{\rm\thinspace K}}
\def\keV{{\rm\thinspace keV}}
\def\km{{\rm\thinspace km}}
\def\kpc{{\rm\thinspace kpc}}

\def\Mpc{{\rm\thinspace Mpc}}
\def\Msun{\hbox{$\rm\thinspace M_{\odot}$}}
\def\pc{{\rm\thinspace pc}}
\def\ph{{\rm\thinspace ph}}
\def\s{{\rm\thinspace s}}

\def\sr{{\rm\thinspace sr}}
\def\Hz{{\rm\thinspace Hz}}

\def\cmps{\hbox{$\cm\s^{-1}\,$}}
\def\cmsq{\hbox{$\cm^2\,$}}

\def\pcmcu{\hbox{$\cm^{-3}\,$}}

\def\ergpcmsqps{\hbox{$\erg\cm^{-2}\s^{-1}\,$}}

\def\ergps{\hbox{$\erg\s^{-1}\,$}}

\def\kmps{\hbox{$\km\s^{-1}\,$}}

\def\pcmsq{\hbox{$\cm^{-2}\,$}}

\def\phpcmsqpspkeV{\hbox{$\ph\cm^{-2}\s^{-1}\keV^{-1}\,$}}
\def\pHz{\hbox{$\Hz^{-1}\,$}}

\def\ps{\hbox{$\s^{-1}\,$}}

\def\psr{\hbox{$\sr^{-1}\,$}}

\def\kmpspMpc{\hbox{$\kmps\Mpc^{-1}$}}

\def\pct{\%}

\documentclass[]{emulateapj}

\usepackage{natbib}
\shorttitle{The Halo, Hot Spots and Jet/Cloud Interaction of PKS 2153--69}

\shortauthors{Young, Wilson, Tingay \& Heinz}

\begin{document}

\title{The Halo, Hot Spots and Jet/Cloud Interaction of PKS 2153--69}

\author{A. J. Young\footnotemark[1], A. S. Wilson\footnotemark[2,3], S. J.
Tingay\footnotemark[4] and S. Heinz\footnotemark[1,5]}

\footnotetext[1]{MIT Center for Space Research, 77 Massachusetts Avenue,
Cambridge, MA 02139}

\footnotetext[2]{Astronomy Department, University of Maryland, College Park, MD
20742}

\footnotetext[3]{Adjunct Astronomer, Space Telescope Science Institute, 3700
San Martin Drive, Baltimore, MD 21218}

\footnotetext[4]{Centre for Astrophysics and Supercomputing, Swinburne
University of Technology, Mail No 31, P.O. Box 218, Hawthorn, Vic. 3122,
Australia}

\footnotetext[5]{Chandra fellow}



\begin{abstract}

We report Chandra X-ray Observatory and 1.4 GHz Australian Long Baseline Array
(LBA) observations of the radio galaxy PKS 2153--69 and its environment.  The
Chandra image reveals a roughly spherical halo of hot gas extending out to
$30\kpc$ around PKS 2153--69.  Two depressions in the surface brightness of the
X-ray halo correspond to the large scale radio lobes, and interpreting these as
cavities inflated with radio plasma we infer a jet power of $4 \times 10^{42}
\ergps$.  Both radio lobes contain hot spots that are detected by Chandra.  In
addition, the southern hot spot is detected in the 1.4~GHz LBA observation,
providing the highest linear resolution image of a radio lobe hot spot to date. 
The northern hot spot was not detected in the LBA observation.  The radio to
X-ray spectra of the hot spots are consistent with a simple power law
emission model.  The nucleus has an X-ray spectrum typical of a type 1 active
galactic nucleus, and the LBA observation shows a one-sided nuclear jet on
$0\farcs1$ scales.  Approximately $10\arcsec$ northeast of the nucleus, X-ray
emission is associated with an extra-nuclear cloud.  The X-ray emission from
the cloud can be divided into two regions, an unresolved western component
associated with a knot of radio emission (in a low resolution map), and a
spatially extended eastern component aligned with the pc-scale jet and
associated with highly ionized optical line-emitting clouds.  The X-ray spectrum of the eastern component is very soft ($\Gamma > 4.0$ for a power law model or $kT \simeq 0.22 \keV$ for a thermal plasma).  The LBA
observation did not detect compact radio emission from the extra-nuclear
cloud.  We discuss both jet precession and jet deflection models to account for
the progressively increasing position angle from the northern hot spot to the
western component of the jet/cloud interaction region to the direction of the
pc-scale jet.  In the precession model the particle beam impacts the western
region while the radiation beamed from the nucleus photoionizes the eastern
region and is scattered into our line of sight by dust giving rise to the polarized optical emission and possibly the soft X-ray emission.  If the X-ray emission from the eastern region really is dust-scattered nuclear radiation, it would be the first detection of such emission from an external galaxy.  The nearby galaxy MRC 2153--699 is also detected by Chandra.

\end{abstract}


\keywords{galaxies : active --- galaxies : individual (PKS 2153--69) ---
galaxies : jets --- magnetic fields --- X-rays : galaxies --- cosmic rays}


%

\section{Introduction}

The morphology of a typical radio galaxy is strongly influenced by interactions
between the relativistic jet that  originates at its nucleus and material in
the interstellar and intergalactic media.  This appears true at all stages of
radio galaxy evolution.  In the GHz Peaked Spectrum and Compact Steep Spectrum
objects, in many cases thought to be radio  galaxies in the early stages of
evolution \citep{2003PASA...20...38S}, interactions between their jets and
their nuclear environment have been observed
\citep[e.g.][]{2002RMxAC..13..196O}.  In the more mature, low-power FR-I type
radio galaxies, entrainment of material by a relativistic jet has been
suggested as the mechanism responsible for their  expanding, decelerating jets
and center-brightened radio lobes \citep[e.g.][]{1994ApJ...422..542B}.  In
contrast, in the  higher-power FR-II radio galaxies, interactions between the
jets and the intergalactic medium terminate the jets in spectacular lobe
hot spots, giving these radio galaxies their characteristic edge-brightened
large-scale morphologies \citep[i.e.][]{1991ApJ...383..554C, 1997A&A...328...12P}.

The jets themselves, and the interaction regions, produce emission via both thermal and  non-thermal mechanisms over a wide range of spatial scales, from sub-pc to Mpc.  Much of the detailed physics pertaining to relativistic jets in radio galaxies is still uncertain \citep{de_young_04} and the complexity of the physics involved in interactions between these jets and their environments is underscored by a number of recent theoretical and numerical simulation studies \citep[e.g.][]{2002ApJ...572..713H}.

With these considerations in mind it is immediately apparent that high spatial
resolution, multi-wavelength  observations of radio galaxies are a prerequisite
for detailed investigations of the physical processes involved. The benefits of
high spatial resolution, multi-wavelength observational studies of radio
galaxies are illustrated by recent work on the nearby radio galaxies Pictor A
\citep{2001ApJ...547..740W, 1997A&A...328...12P} and Centaurus A
\citep{2003ApJ...593..169H}.

PKS 2153--69 is one of the few nearby and powerful radio galaxies, a transition
FR-I/FR-II object \citep{1998MNRAS.296..701F} at a redshift of $z = 0.0282$
\citep{1988MNRAS.235..403T}.  The proximity of PKS 2153--69 ($20\pct$ closer than Pictor A) makes  it an ideal object in which to study a relativistic jet and its relationship to its environment.  As well as the  strong radio lobe
hot spots typical of FR-II radio galaxies, PKS 2153--69 also contains one of the
best examples of  an interaction between a relativistic jet and an
extra-nuclear cloud of gas \citep{1987Natur.325..504T, 1988MNRAS.235..403T,
1998MNRAS.296..701F, 1990iuea.rept..513F, 1988Natur.334..591D}.  Radio and 
optical imaging by \citet{1998MNRAS.296..701F} and VLBI observations by
\citet{1996AJ....111..718T} have provided good evidence that a jet originating
at the nucleus of the host galaxy impacts a cloud of gas in the outskirts of
the galaxy,  possibly the remnant of a merging galaxy.  Near the interaction
site, an optically emitting cloud (line and continuum) with a complex 
morphology and ionization structure is seen.  

In this paper we present observations that address the twin goals of
multi-wavelength coverage and improved spatial resolution.  First, comprehensive
Chandra X-ray imaging observations of PKS 2153--69 are presented,  including
analyses of the major X-ray emitting regions of the system: the hot halo, the
nucleus, the jet/cloud  interaction, and the lobe hot spots.  Second, high spatial resolution radio observations (VLBI) are presented  that are
targeted to the aforementioned regions of interest.  In particular, we present
the highest linear resolution observation of a radio galaxy lobe hot spot yet
produced.

We attempt to produce a coherent physical picture of the jet interaction
regions in this radio galaxy by considering the Chandra X-ray data, the VLBI
data, archival HST data, and other published data.

We use the WMAP measured cosmology of $H_0 = 71 \kmpspMpc$ and $\Omega_\Lambda
= 0.73$ \citep{2003ApJS..148....1B}.  PKS 2153--69 has a redshift of $z =
0.0282$ which corresponds to a luminosity distance of $123 \Mpc$ and $1\arcsec
= 566 \pc$.  The Galactic column density towards PKS 2153--69 is $N_{\rm H} =
2.5 \times 10^{20} \pcmsq$ \citep{1990ARA&A..28..215D}.

\section{Observations}

\subsection{X-ray}

PKS 2153--69 was observed by the Chandra X-ray observatory on 2001-08-02 for
14~ks.  The nucleus was placed at the aim-point of the ACIS-S detector, and
CCDs 2, 3, 5, 6, 7 and 8 were active, in the standard full-frame mode with a
frame-time of $3.2\s$.  The observation was mildly affected by background
flares that increased the background count rate to approximately 4/3 the
quiescent rate.  Since most of the X-ray emitting regions are small we chose
not to remove the periods of high background since this would significantly
degrade the signal-to-noise ratio. Data were extracted using CIAO 3.0.2 with
CALDB 2.26, and analyzed using XSPEC 11.3.0.  The degradation of the low energy
response of ACIS due to molecular contamination is taken into account using the
latest CIAO tools.  The Chandra astrometry is excellent and the position of the X-ray nucleus differs from the position of the radio nucleus quoted by \citet{1998MNRAS.296..701F} by only $0\farcs2$.  When overlaying the radio maps of \citet{1998MNRAS.296..701F} on our Chandra images we have aligned the brightest pixel in the radio map with the brightest pixel in the Chandra image.

\subsection{Radio}

PKS 2153--69 was observed with three elements of the Australian
Long Baseline Array  (LBA) on 2003 February 15: the 22 m Mopra telescope, three
of the six 22 m antennas of the Australia  Telescope Compact Array (ATCA) tied
together as a phased array, and the 64 m Parkes telescope.  The observation was
conducted over a 6 hour period, the data recorded using the S2 system
\citep{1996ITIM.45(6).923W} consisting of dual polarization 16 MHz bands
centered at 1400 MHz.  Each 16 MHz band  was 2-bit sampled, giving an aggregate
recorded data rate of 128 Mbps.

The recorded data were shipped to the LBA correlator
\citep{1996.....conf...16W} and correlated using a 5 second integration time
and 32 frequency channels across each 16 MHz band.  The cross polarization
products were not correlated.  The synthesized beam, using uniform weighting,
gives a restoring beam for imaging of approximately 90 mas $\times$ 150 mas at
a position angle of $-67\degmark$.

Once correlated, system temperatures and antenna gains were applied to the data
in AIPS\footnote{The Astronomical Image Processing Software (AIPS) has been
developed and is  maintained by the National Radio Astronomy Observatory, which
is operated by Associated Universities, Inc., under cooperative agreement with
the National Science Foundation}, to  calibrate the visibility amplitudes. 
Adjustments to the calibration were derived from a short observation of a
strong unresolved source (PKS B1921$-$293) and  applied to the PKS 2153--69
data.  The data were then fringe-fitted in AIPS, each polarization
independently.  The residual delays and rates were applied to the data before
the data were exported to disk.  The data were not averaged in time or
frequency and all 32 of the 0.5 MHz frequency channels were retained and multi-frequency synthesis techniques are used in the imaging process.

The strong and compact source at the nucleus of the radio galaxy (the phase
tracking center used at the correlator) was imaged in DIFMAP
\citep{1997adass...6...77S}, following editing of the data and averaging in
time over a 10 second timescale.  With only three antennas in the  array only
phase self-calibration was possible.

The self-calibrated data following imaging of the nuclear radio source were
imported into the MIRIAD processing software \citep{1995adass...4..433S}.  In
MIRIAD, a model of the  nuclear source was produced in the ($u,v$) plane and
was subtracted from the data.  Then the data were edited so that the
phase-tracking center was shifted, entailing an appropriate re-calculation of
($u,v$) coordinates.  Three  shifted datasets were produced in this way.  The
first of the shifted datasets placed the phase-tracking center on the brightest
part of the southern lobe  of the radio galaxy (the southern lobe  hot spot). 
The second shifted dataset placed the phase-tracking center at the northern
lobe hot spot. Finally, the third shifted dataset was centered at the location
that marks the point of interaction between the northern jet and a cloud of gas
in the host galaxy.  The coordinates of these three new  phase-tracking centers
were estimated from the published ATCA images of \cite{1998MNRAS.296..701F}. 
In this way, images of these three regions could be formed by Fourier inversion
and deconvolution.

\subsection{Archived Hubble Space Telescope Image}

A Hubble Space Telescope (HST) F606W image of PKS 2153--69 was obtained from the HST archive.  This is the same image as presented in the optical/radio comparison of PKS 2153--69 in \citet{1998MNRAS.296..701F}.  The HST image was initially cleaned of cosmic rays using the Figaro task bclean.  The coordinate system of the HST image was then shifted to align the brightest pixel ($\alpha = 21^{\rm h} 57^{\rm m} 06\fs248$; $\delta = -69\arcdeg 41\arcmin 23\farcs79$) with the brightest radio and X-ray pixels ($\alpha = 21^{\rm h} 57^{\rm m} 06\fs038$; $\delta = -69\arcdeg 41\arcmin 23\farcs78$).

\section{Morphology}

\subsection{Hot Halo and Radio Lobes}

\begin{figure}
\centerline{\includegraphics[scale=0.4,angle=270]{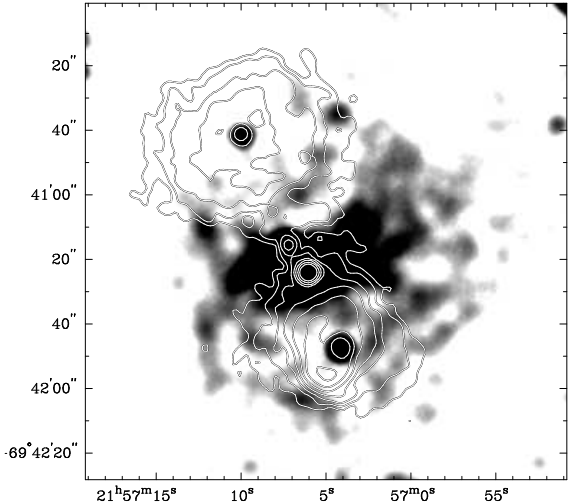}}

\caption{Chandra X-ray image in the 0.5 -- 7 keV band, with contours of an ATCA
4.7~GHz radio map \citep{1998MNRAS.296..701F} overlaid.  Here and throughout the paper, coordinates are J2000.0 epoch.  The Chandra image has been smoothed by a Gaussian of FWHM $6\arcsec$ and the radio map has been smoothed by a Gaussian of FWHM $3\arcsec$.  A diffuse halo of hot gas is seen in the X-ray image, extending out to a radius comparable to that of the radio lobes.  There are depressions in the X-ray surface brightness at the locations of the radio lobes.  Each radio lobe contains a radio and X-ray hot spot. \label{fig:cluster}}

\end{figure}

\begin{figure}
\centerline{
  \includegraphics[scale=0.4,angle=270]{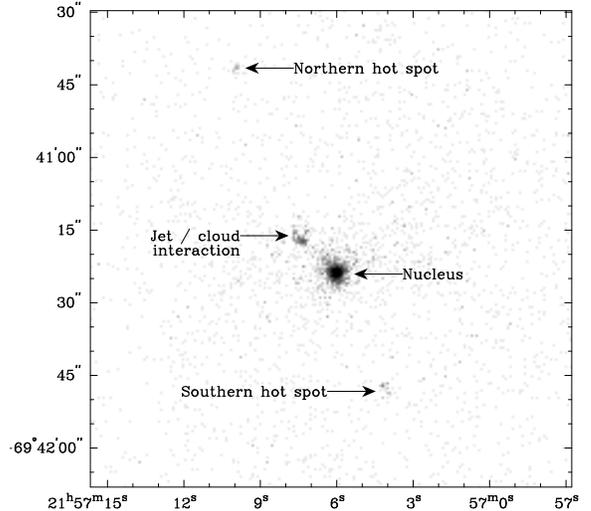}
}

\caption{Full resolution Chandra X-ray image in the 0.5 -- 7 keV band with interesting regions labeled. \label{fig:cxo_large_hires}}

\end{figure}

The large scale 0.5 -- 7 keV X-ray and 4.7 GHz radio morphologies of PKS
2153--69 are shown in Fig.~\ref{fig:cluster}.  The radio map shows two lobes, each of which contains a hot spot. 
The X-ray image shows a halo of hot gas extending out to a radius comparable to
that of the radio lobes, approximately $1\arcmin = 30\kpc$.  There are
depressions in the X-ray surface brightness at the location of the radio lobes
suggesting that the radio plasma has inflated cavities in the inter-galactic
medium.  The southern cavity is completely embedded in the halo gas but the northern cavity appears to have ``broken out'' of the halo to the northeast and may be less well confined by the halo gas.  A higher resolution X-ray image is shown in Fig.~\ref{fig:cxo_large_hires}, and the highlighted regions are discussed below.

\subsection{Nucleus} \label{sec:morph_nuc}

\begin{figure}
\centerline{\includegraphics[scale=0.4,angle=270]{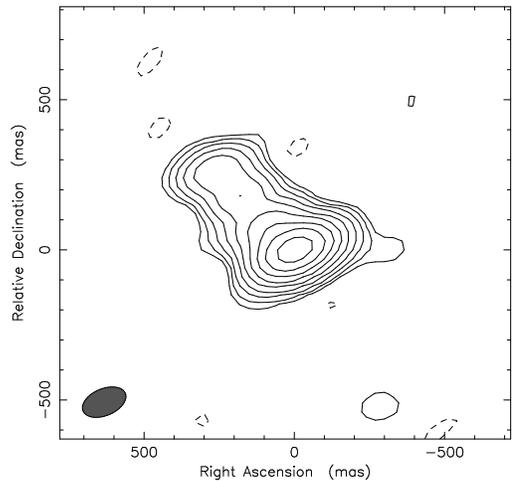}}

\caption{1.4 GHz VLBI image of the radio nucleus of PKS 2153--69, as described
in the text.  Contours are $-0.25, 0.25, 0.5, 1, 2, 4, 8, 16, 32$ and $64\pct$ of the peak, 0.379 Jy/beam.  The restoring beam size is $152 \times 90$ mas at a major axis position angle of $-67\degmark$. \label{fig:vlbi_nuc}}

\end{figure}

The VLBI observation (Fig.~\ref{fig:vlbi_nuc}) shows that the nuclear radio
source consists of a strong unresolved component (the core) and a weaker
extension along position angle $\sim 45\degmark$ (the jet).  Assuming the jet
is continuous and has a spectral index of $\alpha = 0.7$ ($S \propto \nu^{-\alpha}$), the flux ratio between the jet and counter jet is:

\begin{equation}
R = [(1+\beta_{\rm j}\cos\theta)/(1-\beta_{\rm j}\cos\theta)]^{2+\alpha} > 8.1
\end{equation}

from which we derive:

\begin{equation} \label{eqn:j_cj}
\beta_{\rm j} \cos \theta > 0.37
\end{equation}

for the pc scale jet, where $\beta_{\rm j} = v_{\rm j} / c$ is the velocity of
the material in the jet and $\theta$ the angle between the jet and the line of sight.

\begin{figure}
\centerline{\includegraphics[scale=0.4,angle=270]{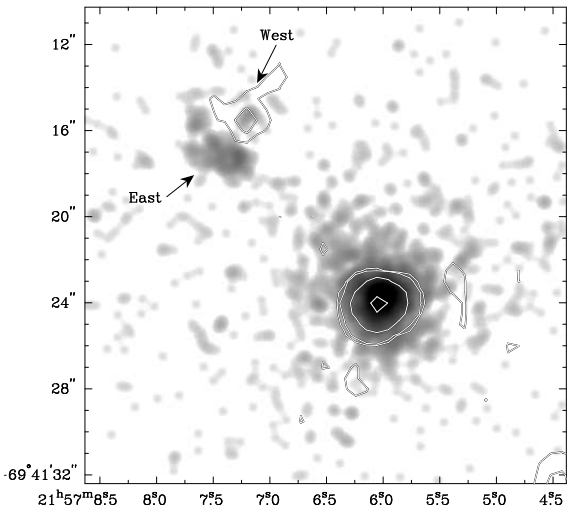}}

\caption{Chandra 0.5 -- 7 keV X-ray image of the nucleus and jet/cloud
interaction region of PKS 2153--69 with contours of an ATCA 8.6 GHz radio map
\citep{1998MNRAS.296..701F} overlaid.  The Chandra image has been resampled to
ten times smaller pixel size and smoothed by a Gaussian of FWHM $0\farcs5$. 
The X-ray emission from the jet/cloud interaction region approximately $10\arcsec$ northeast of the nucleus can be divided into two
regions, a spatially extended eastern region (labeled ``East'') and an
unresolved western region (labeled ``West'').  The western region is associated
with a knot of radio emission. \label{fig:cxo_nuc_hic} }

\end{figure}

The X-ray nucleus is also spatially extended toward the northeast on a much
larger scale (Fig.~\ref{fig:cxo_nuc_hic}).  A more quantitative description of
the X-ray morphology of the nucleus is difficult because the ACIS image suffers
from pile-up, having an observed count rate of $0.17$~cts~$\ps$.

\subsection{Jet/Cloud Interaction Region}

\begin{figure*}
\centerline{\includegraphics[scale=1.0]{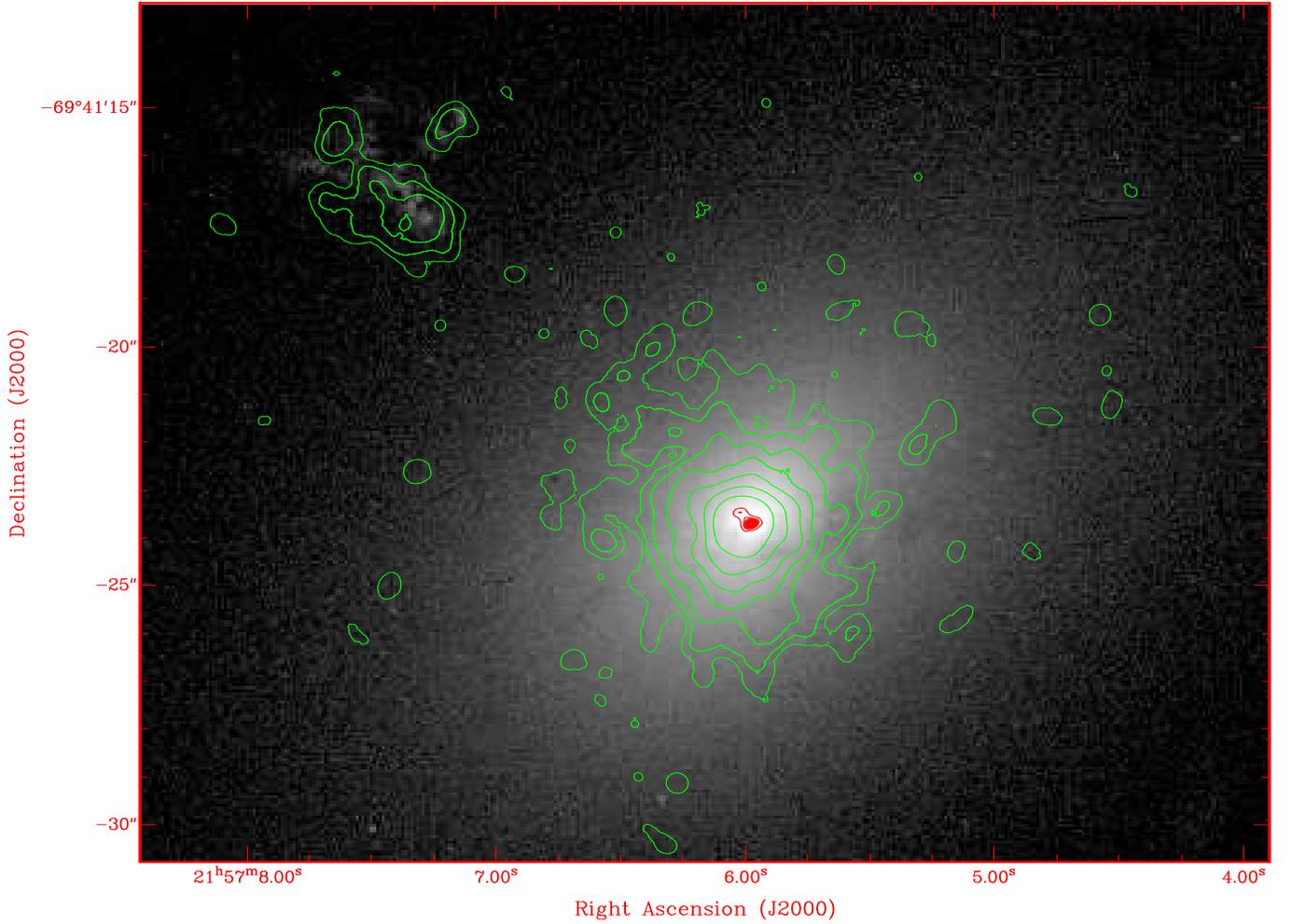}}

\caption{HST F606W image of the PKS 2153--69 host galaxy (grey scale) with
Chandra X-ray contours (green) overlaid.  Also, 1.4 GHz radio contours for the
nuclear radio source (red) are overlaid. \label{fig:cxo_hst_vlbi}}

\end{figure*}

Approximately $10\arcsec$ northeast of the nucleus there are a series of clouds
that lie roughly along the position angle of the VLBI jet.  The X-ray emission from the clouds can be divided into two regions (Fig.~\ref{fig:cxo_nuc_hic}).  Firstly,
centered at a nuclear distance of $9\farcs0$ along P.A. $45\degmark$ there is a
region of X-ray emission that is spatially resolved, with an extent of $\simeq
4\farcs5$ along the jet direction and an extent of $\simeq 2\farcs5$ transverse
to the jet direction.  We shall refer to this region as the eastern component
of the jet/cloud interaction region.  Secondly, at a nuclear distance of
$10\farcs6$ along P.A. $34\degmark$ (Fig.~\ref{fig:cxo_nuc_hic}) there is a
compact source of X-ray emission.  We shall refer to this region as the
western component of the jet/cloud interaction region.  To test whether the western region is resolved by Chandra we consider the counts within a $0\farcs5$ radius circle and a surrounding annulus of inner radius $0\farcs5$ and outer radius $2\farcs5$.  The region of the annulus that overlaps the eastern component is excluded, and the counts in the remainder of the annulus are scaled to account for the lost area.  The background count rate due to the diffuse halo is estimated from a region at a similar nuclear distance to the east of the nucleus.  We find that $75\pct$ of the counts above the diffuse background level come from within $\approxlt 0\farcs5$ of the western region.  This is comparable to that expected for a soft X-ray point source ($\simeq 80\pct$), so the western component is consistent with being an unresolved point source superimposed on diffuse halo emission.  An optical image of the nucleus and jet/cloud interaction region taken with HST is shown in Fig.~\ref{fig:cxo_hst_vlbi}, with contours of Chandra X-ray emission and VLBI 1.4 GHz radio emission overlaid.  The HST image \citep[Fig.~\ref{fig:cxo_hst_vlbi};][]{1998MNRAS.296..701F} shows optical emission from a number of filamentary clouds in the region associated with the X-ray emission.  Ground-based optical spectra of these filaments show emission lines from a range of very highly ionized ions including strong [Fe X] $\lambda 6375$ \citep{1987Natur.325..504T, 1988MNRAS.235..403T, 1988Natur.334..591D, 1990iuea.rept..513F, 1998MNRAS.296..701F}.  \citet{1998MNRAS.296..701F} detect radio emission from the western component associated with the unresolved X-ray emission (Fig.~\ref{fig:cxo_nuc_hic}).  Our VLBI observation did not detect any radio emission from the jet/cloud interaction region with a $3\sigma$ upper limit of 3 mJy/beam.

\subsection{Southern Hot Spot}

\begin{figure}
\centerline{\includegraphics[scale=0.4,angle=270]{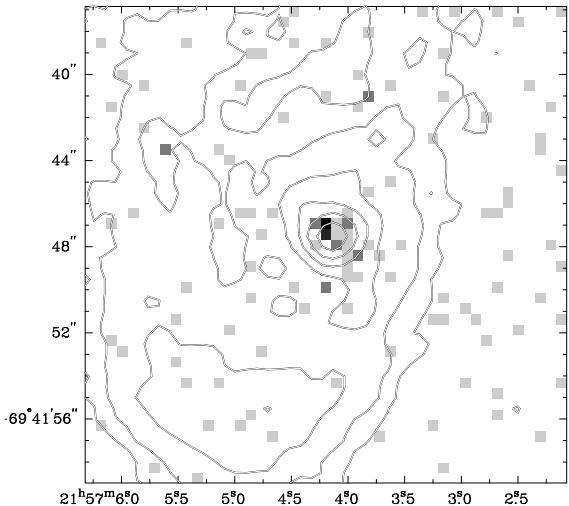}}

\caption{Chandra 0.5 -- 7 keV X-ray image of the southern hot spot with
contours of an 8.6 GHz radio map \citep{1998MNRAS.296..701F} overlaid.
\label{fig:cxo_shs}}

\end{figure}

\begin{figure}
\centerline{\includegraphics[scale=0.4,angle=270]{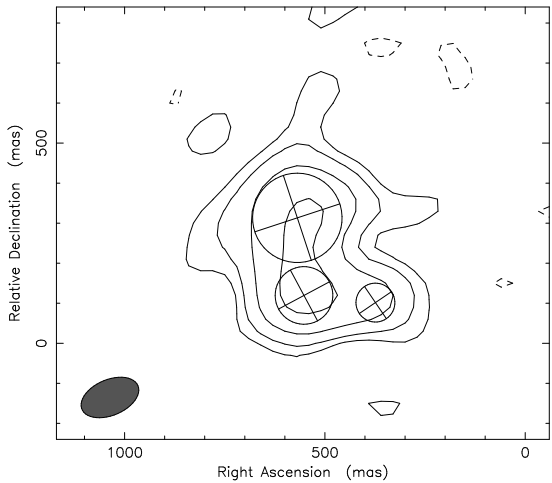}}

\caption{1.4 GHz VLBI image of the southern lobe hot spot of PKS 2153--69, as
described in the text.  Contours are $-10, 10, 20, 40$ and $80\%$ of the peak,
17.2 mJy/beam.  The restoring beam size is $152 \times 90$ mas at a major
axis position angle of $-67\degmark$.  The circles represent the FWHM of the three circular Gaussian components used to model the structure of the radio source (the crosses inside the circles have arbitrary position angles). \label{fig:vlbi_south}}

\end{figure}

At a nuclear distance of $25\arcsec$ along P.A. $202\degmark$
(Fig.~\ref{fig:cxo_shs}) X-ray emission is seen to approximately coincide with
the radio emission from the southern hot spot, although there is a tentative
suggestion that the peak of the X-ray emission is slightly closer to the
nucleus than the peak of the radio emission (the offset is only $\simeq 0\farcs5 \simeq 1$ Chandra pixel and is at the limit of Chandra's astrometric accuracy). The VLBI observation detected
significant emission from the southern lobe hot spot
(Fig.~\ref{fig:vlbi_south}).  The compact radio emission at the southern lobe 
hot spot can be adequately represented in the image plane by three circular
Gaussian components, separated by approximately 400 mas, one component with a
FWHM of 140 mas and a total flux density of 35 mJy, the second component with a
FWHM of 220 mas and a total flux density of 65 mJy, and the third component
with a FWHM of 100 mas and a total flux density of 10 mJy.  The VLBI image of
the southern lobe hot spot shown in Fig. \ref{fig:vlbi_south} is the highest
linear resolution image of a radio galaxy lobe hot spot ever produced, to the
best of our knowledge.  Previously \citet{1997A&A...328...12P} imaged the
western hot spot of Pictor A with the VLA at a wavelength of 2~cm, giving a
best angular resolution of 90~mas.  The best angular resolution of
Fig.~\ref{fig:vlbi_south} is also 90~mas (at a wavelength of 20~cm however),
giving a linear resolution approximately $20\pct$ better than the Pictor A hot
spot since PKS 2153--69 is approximately $20\pct$ closer than Pictor A.

\subsection{Northern Hot Spot} \label{sec:morph:nhs}

\begin{figure}

\centerline{\includegraphics[scale=0.4,angle=270]{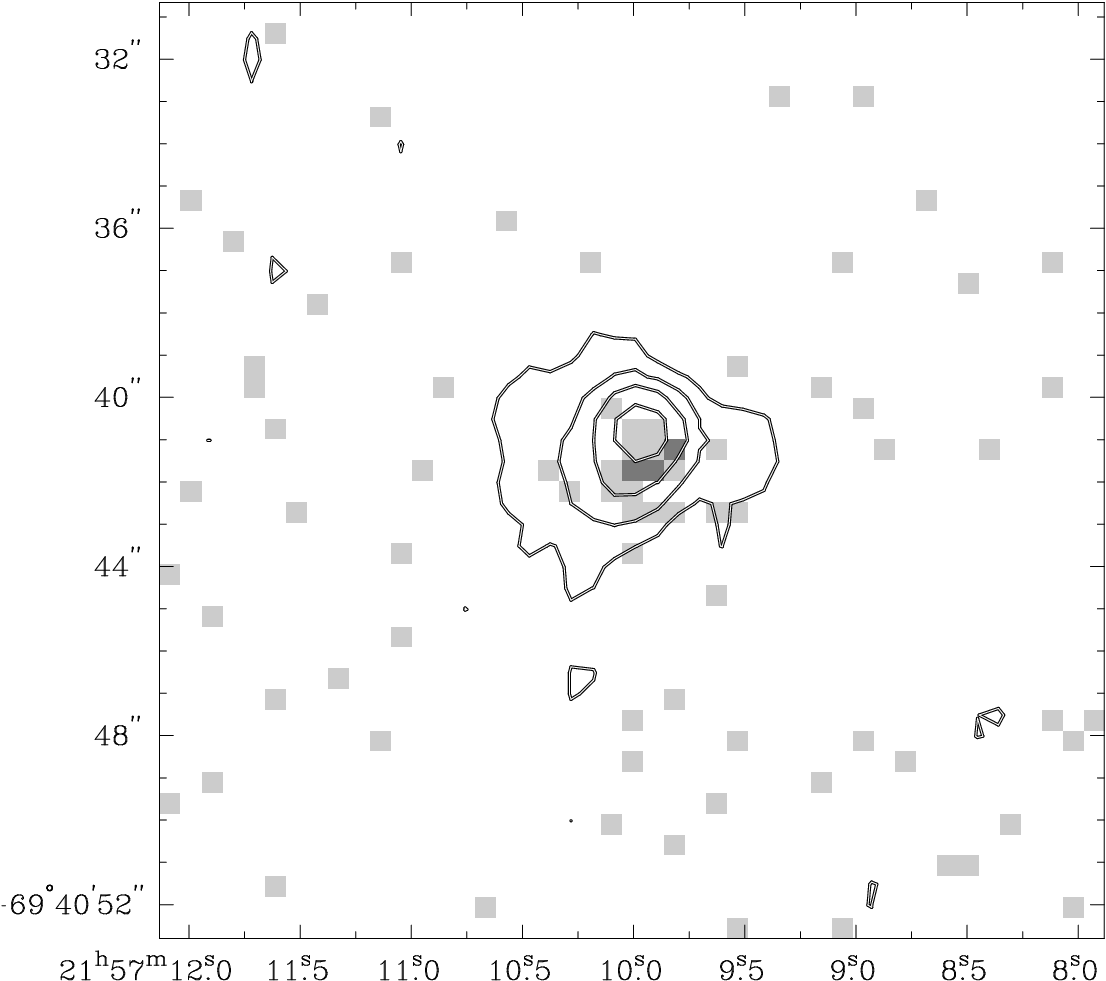}}

\caption{Chandra 0.5 -- 7 keV X-ray image of the northern hot spot with
contours of an 8.6 GHz radio map \citep{1998MNRAS.296..701F} overlaid.
\label{fig:cxo_nhs}}

\end{figure}

At a nuclear distance of $47\arcsec$ along P.A. $26\degmark$
(Fig.~\ref{fig:cxo_nhs}) faint X-ray emission is seen to approximately coincide
with the radio emission from the northern hot spot, although there is a
tentative suggestion that the peak of the X-ray emission is slightly closer to
the nucleus than the peak of the radio emission, as was the case for the
southern hot spot (and also at the limit of Chandra's astrometric accuracy).  Our VLBI observation did not detect any radio emission from
this region with a $3\sigma$ upper limit of 3 mJy/beam.  If we assume that
the northern lobe hot spot has a similar size and structure as the southern
lobe hot spot, we estimate that an upper limit to the total flux density of the
compact emission from the northern lobe hot spot is approximately 25 mJy.

\section{X-ray Spectra}

The extraction regions for the sources and backgrounds are given in Table~\ref{tab:ex}.  Any emission which is unrelated to the region of interest was excluded.

\begin{deluxetable*}{ccccccc}
\tablecaption{Extraction regions for spectra}
\tablecolumns{7}
\tablehead{
    \colhead{Region} & \colhead{Aperture for} & \colhead{Radius/radii} & \colhead{Range of P.A.} & \colhead{Aperture for} & \colhead{Radius/radii} & \colhead{Range of P.A.} \\
    \colhead{} & \colhead{source\tablenotemark{a}} & \colhead{of source} & \colhead{for source} & \colhead{background\tablenotemark{a}} & \colhead{of background} & \colhead{for background} \\
    \colhead{} & \colhead{} & \colhead{aperture} & \colhead{aperture} & \colhead{} & \colhead{aperture} & \colhead{aperture}}
\startdata
Halo & Circle\tablenotemark{b} & $1\farcm1$ & All & Annulus\tablenotemark{c} & Inner -- $1\farcm1$ & All \\
(\S \ref{sec:spec:halo}) & & & & & Outer -- $1\farcm8$ & \\
\\
Nucleus & Circle & $4\farcs4$ & All & Annulus & Inner -- $4\farcs4$ & All \\
(\S \ref{sec:spec:nuc}) & & & & & Outer -- $7\farcs9$ & \\
\\
Jet/Cloud & Annulus\tablenotemark{d} & Inner -- $7\farcs4$ & $41\degmark - 60\degmark$ & Annulus\tablenotemark{d} & Inner -- $7\farcs4$ & $60\degmark - 26\degmark$ \\
Interaction Region -- & & Outer -- $13\farcs3$ & & & Outer -- $13\farcs3$ & \\
Eastern (\S \ref{sec:hic_spec}) & & & & & & \\
\\
Jet/Cloud & Annulus\tablenotemark{e} & Inner -- $7\farcs4$ & $26\degmark - 41\degmark$ & Annulus\tablenotemark{e} & Inner -- $7\farcs4$ & $60\degmark - 26\degmark$ \\
Region -- & & Outer -- $13\farcs3$ & & & Outer -- $13\farcs3$ & \\
Western (\S \ref{sec:hic_spec}) & & & & & & \\
\enddata
\tablenotetext{a}{All apertures are centered on the nucleus.}
\tablenotetext{b}{Nuclues, jet/cloud interaction region and hot spots omitted.}
\tablenotetext{c}{Two bright point sources omitted.}
\tablenotetext{d}{Does not include the unresolved western component of the jet/cloud interaction region.}
\tablenotetext{e}{Does not include the eastern component of the jet/cloud interaction region.}
\label{tab:ex}
\end{deluxetable*}

\subsection{Hot Halo} \label{sec:spec:halo}

The spectrum is well described by a MEKAL \citep{1985A&AS...62..197M, 1986A&AS...65..511M, 1992kaastra, 1995ApJ...438L.115L} thermal plasma model with a temperature of $kT = 1.01\keV$ (see Table~\ref{tab:spec}).  The metallicity is
significantly sub-solar with $Z = 0.28 Z_\odot$.

\subsection{Nucleus} \label{sec:spec:nuc}

The nuclear spectrum suffers from the effects of pile-up which are
corrected using the routine of \citet{2001ApJ...562..575D} that is available in
XSPEC \citep{1996adass...5...17A}.  The spectrum is well described by a power law model with photon index $\Gamma = 1.75$ and a 2 -- $10\keV$ luminosity of $8.7 \times 10^{42}\ergps$ absorbed by the Galactic column (see Table~\ref{tab:spec}).  This is typical of a ``type 1'' active galaxy or quasar.           

\begin{deluxetable*}{ccccccccc}

\tablecaption{X-ray spectral models.\tablenotemark{1}}

\tablecolumns{9}

\tablehead{\colhead{Region} & \colhead{Model} & \colhead{Parameter} &
\colhead{Normalization}\tablenotemark{2} & \multicolumn{2}{c}{Unabsorbed flux}
& \multicolumn{2}{c}{Unabsorbed luminosity} & \colhead{$\chi^2$/dof} \\ & & & & \multicolumn{2}{c}{[$\ergpcmsqps$]} & \multicolumn{2}{c}{[$\ergps$]} &  \\ & & & & \colhead{0.5 -- $2\keV$} & \colhead{2 -- $10\keV$} & \colhead{0.5 -- $2\keV$} & \colhead{2 -- $10\keV$} &}

\startdata

Halo & MEKAL & $kT = 1.01^{+0.06}_{-0.06} \keV$ & $5.4^{+1.3}_{-0.9} \times
10^{-4}$ & $3.9 \times 10^{-13}$ & & $6.8 \times 10^{41}$ & & \\

& metallicity & $Z = 0.28^{+0.10}_{-0.09} Z_\odot$ & & & & & & 44.4/60 \\

Nucleus & PL & $\Gamma = 1.75^{+0.08}_{-0.05}$ & $1.4^{+0.1}_{-0.7} \times
10^{-3}$ & $3.0 \times 10^{-12}$ & $5.0 \times 10^{-12}$ & $5.1 \times 10^{42}$
& $8.7 \times 10^{42}$ & 117.7/129 \\

Jet/cloud (E) & MEKAL & $kT = 0.22^{+0.02}_{-0.03} \keV$ & $2.5^{+0.7}_{-0.5}
\times 10^{-5}$ & $3.4 \times 10^{-14}$ & & $6.2 \times 10^{40}$ & & \\

& plus PL & $\Gamma = 1.0$\tablenotemark{3} & $2.6^{+0.8}_{-1.6} \times
10^{-6}$ & $6.1 \times 10^{-15}$ & & $1.0 \times 10^{40}$ & & 3.5/5 \\

\enddata

\tablenotetext{1}{All models include absorption by only the Galactic column.  The
error bars indicate the 90 per cent confidence interval for a single
interesting parameter ($\Delta \chi^2 = 2.7$).}

\tablenotetext{2}{The power law normalization is $\phpcmsqpspkeV$ at $1\keV$. 
The MEKAL thermal plasma model normalization is $ 10^{-14} \int n_e n_H dV/(4
\pi [D_A(1+z)]^2) $ where $D_A$ is the angular size distance ($\cm$), and $n_e$
and $n_H$ are the electron and hydrogen densities ($\pcmcu$), respectively.}

\tablenotetext{3}{Parameter fixed.}

\label{tab:spec}
\end{deluxetable*}

\subsection{Jet/Cloud Interaction Region} \label{sec:hic_spec}

\begin{figure}
\centerline{\includegraphics[scale=0.4,angle=270]{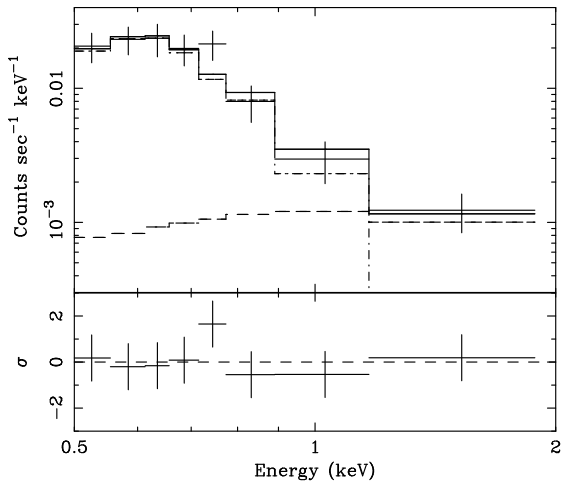}}

\caption{Chandra X-ray spectrum of the eastern component of the jet/cloud
interaction region.  The upper panel shows the spectrum (`$+$' symbols) with
the best fit model overlaid (solid line).  The model consists of a thermal
plasma component (dot-dashed line) and a hard power law component (dashed
line).  The lower panel shows the $\chi$ residuals to the model.  See
Section~\ref{sec:hic_spec} and Table~\ref{tab:spec} for details.
\label{fig:hic_fit}}

\end{figure}

If the
spectrum of the extended eastern component is modeled as a power law absorbed by the Galactic column we find that
the spectrum is very soft with $\Gamma > 4.0$, although this gives a poor fit
with a $\chi^2$/dof of 10.5/6.  A thermal plasma of solar abundance
absorbed by the Galactic column provides a slightly better description of the
spectrum with $kT = 0.22\keV$ and a $\chi^2$/dof of 9.8/6.  The poor
$\chi^2$ value of the thermal plasma model is a result of the spectrum being
slightly harder than the model, with excess flux around $1.5\keV$ (there are
too few counts to model the spectrum above $\sim 2\keV$).  An excellent
description of the spectrum is obtained if a power law of photon index $\Gamma
= 1$ is added to the thermal plasma model, giving a $\chi^2$/dof of 3.5/5
(see Table~\ref{tab:spec} and Fig.~\ref{fig:hic_fit}).  The power law component
accounts for approximately 18 per cent of the unabsorbed 0.5 -- $2\keV$ flux. 
We do not have sufficient S/N to strongly constrain this hard component, but a
power law of photon index $\Gamma = 1$ is equivalent to a very hot
bremsstrahlung component.  Thus, an equally good description of the spectrum is
obtained by replacing the power law component with a second, high temperature
thermal plasma ($kT > 3.9\keV$).  Attempting to model the spectrum as a sum of
two power law components absorbed by the Galactic column gives an unacceptable
$\chi^2$/dof of 10.5/4 with significant systematic residuals below
$0.8\keV$ where the thermal plasma model has strong iron L shell emission
lines.  We conclude that the X-ray emission from the extended eastern component
of the jet/cloud interaction region may originate from a thermal plasma
with $kT = 0.22\keV$, with a weaker high temperature or non-thermal component.  An alternative interpretation, discussed in Section \ref{sec:dust}, is that the X-rays from this eastern component are dust-scattered radiation of a blazar X-ray beam from the nucleus.  This beam would be directed at the eastern part of the cloud.  However, we continue here our discussion of the thermal plasma interpretation.

Assuming the eastern component arises from shock heating of a cold gas cloud we can estimate the shock velocity.  The Rankine-Hugoniot shock jump conditions \citep{1959flme.book.....L} give a shock velocity $v_s$ of

\begin{equation}
v_s^2 = \frac{16kT}{3\mu m_H} \label{eq:shock}
\end{equation}

where $\mu = 0.61$ is the mean mass per particle in units of the proton mass, and we have assumed a fully ionized gas and solar abundances.  For a post-shock temperature of $kT = 0.22 \keV$ we estimate a shock velocity of $v_s \simeq 430 \kmps$. 

The unresolved western component of the jet/cloud interaction region is too
faint to obtain its X-ray spectrum directly, but we can form a hardness ratio
$R$ based on the number of counts $S$ in a soft band ($0.5 - 1.2\keV$), and the
number of counts $H$ in a hard band ($1.2 - 5\keV$), given by $R = (H-S) /
(H+S)$.  We find $R = -0.13^{+0.23}_{-0.22}$. 
If we assume the spectrum is a power law absorbed by the Galactic column this
corresponds to $\Gamma = 2.1^{+0.5}_{-0.5}$.  The error bars on $R$ and
$\Gamma$ quoted here (and in subsequent sections dealing with hardness ratios)
denote the 90 per cent confidence interval assuming independent Poisson
statistics for $S$ and $H$.  If the spectrum is a power law of photon index
$\Gamma = 2.1$ absorbed by the Galactic column the flux density is $2.3 \times
10^{-32}\ergpcmsqps\pHz$ at $1\keV$ (see Table~\ref{tab:flux}).  If, on the
other hand, the spectrum is a thermal plasma this hardness ratio corresponds to
a temperature of $kT = 2.5^{+3.8}_{-1.0} \keV$, with the same flux density.
\label{sec:rso_hr}

\begin{deluxetable}{ccc}
\tablecaption{X-ray flux densities.}
\tablecolumns{3}

\tablehead{\colhead{Region} & \colhead{$\log_{10 }({\rm frequency})$} &
\colhead{$\log_{10}({\rm flux\ density})$} \\ & \colhead{(Hz)} &
\colhead{($\ergpcmsqps\pHz$)}}

\startdata

Jet/cloud (west) & 17.4 & $-31.6$ \\

Northern hot spot & 17.4 & $-31.9$ \\

Southern hot spot & 17.4 & $-31.7$ \\

\enddata
\label{tab:flux}
\end{deluxetable}

\subsection{Northern and Southern Hot Spots}

Neither hot spot has sufficient counts to obtain an X-ray spectrum.  We can,
however, determine their hardness ratios using the technique and hard and soft bands described in Section~\ref{sec:rso_hr}.  We find that the northern hot spot has $R = -0.16^{+0.29}_{-0.25}$ which, for a power law model absorbed by the Galactic column, corresponds to $\Gamma = 2.2^{+0.6}_{-0.7}$.  The flux density at 1~keV assuming such a spectrum is $1.4 \times 10^{-32} \ergpcmsqps\pHz$ (see Table~\ref{tab:flux}).  Similarly, we find that the southern hot spot has $R = -0.19^{+0.25}_{-0.21}$ which, for a power law model absorbed by the Galactic column corresponds to $\Gamma = 2.2^{+0.6}_{-0.5}$. The flux density assuming such a spectrum is $1.9 \times 10^{-32} \ergpcmsqps\pHz$ (see Table~\ref{tab:flux}).

\subsection{The Nearby Galaxy MRC 2153--699}

\begin{figure}

\centerline{\includegraphics[scale=0.4,angle=270]{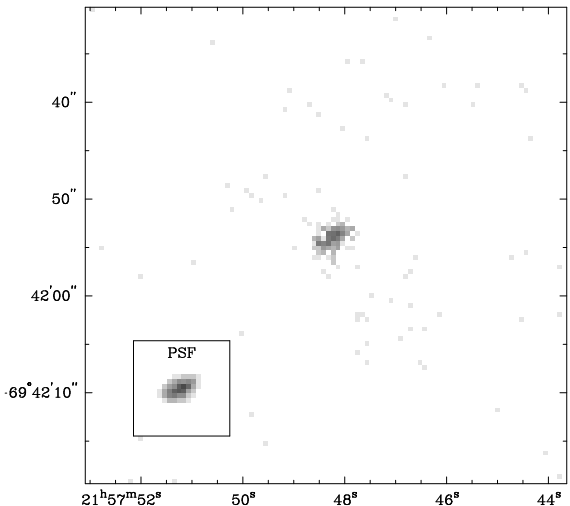}}

\caption{Chandra 0.5 -- 7 keV X-ray image of the radio galaxy MRC 2153--699
that lies almost $4\arcmin$ to the east of PKS 2153--69.  The off-axis point
spread function (PSF) of Chandra is shown in the inset panel, and the nucleus
is not clearly resolved. \label{fig:mrc2153-699}}

\end{figure}

We also detect X-ray emission from the nucleus of the radio galaxy MRC
2153--699 that lies almost $4\arcmin$ to the east of PKS 2153--69 (Fig.~\ref{fig:mrc2153-699}).  The spectrum of MRC 2153--699 is
reasonably well described ($\chi^2/{\rm d.o.f} = 10 /8$) by an absorbed
($N_{\rm H} = 1.1^{+1.0}_{-1.0} \times 10^{22} \pcmsq$) power law ($\Gamma =
0.9^{+0.7}_{-0.4}$) with an observed 1 -- 5 keV X-ray flux of $1.9 \times 10^{-13} \ergpcmsqps$.  Its redshift and therefore luminosity are unknown, although it has been identified with a $20^{\rm th}$ magnitude galaxy \citep{1992ApJS...80..137J}, and imaged in the radio by \citet{1998MNRAS.296..701F}.

\section{Discussion}

\subsection{Hot Halo} \label{sec:dis_halo}

The hot halo has a radius of approximately $1\arcmin$, a temperature of $T
\simeq 10^7\K$, an electron number density of $n_{\rm e} \simeq 7 \times
10^{-3} \pcmcu$ and a pressure $P \simeq 10^{-11}$ dyn cm$^{-2}$.  The electron density is calculated using the emission measure of the X-ray thermal plasma model and assuming the halo is a uniform density spherical cloud.  The total mass of halo gas is $M_{\rm gas} \simeq 3 \times 10^{10} \Msun$, which is probably only a small fraction of the gravitating mass in the halo.  If we assume that the depressions in the X-ray emitting halo corresponding to the radio lobes are cavities filled with relativistic radio plasma we can estimate the power of the jet.  The sound speed in the halo is $c_{\rm s} \simeq 3 \times 10^7 \cmps$.  The southern lobe has a radius of $\simeq 15\farcs5$, and the southern hot spot is $25\arcsec$ from the nucleus.  Using the sound crossing time, $t_{\rm sc}$, from the nucleus to the southern hot spot as a characteristic timescale we estimate the jet power required to inflate the southern cavity to be $F_{\rm j} \simeq 4 PV_{\rm cavity} / t_{\rm sc} = 2 \times 10^{42} \ergps$, where $V_{\rm cavity}$ is the volume of the cavity.  Since there are two cavities the total jet power is $F_{\rm j} \simeq 4 \times 10^{42} \ergps$, which is a factor of $\sim 3$ below the X-ray luminosity of the nucleus (in the 0.5 -- 10 keV band).

\subsection{Northern and Southern Hot Spots} \label{sec:discussion:n_s_hs}

\begin{figure}
\centerline{\includegraphics[scale=0.4,angle=270]{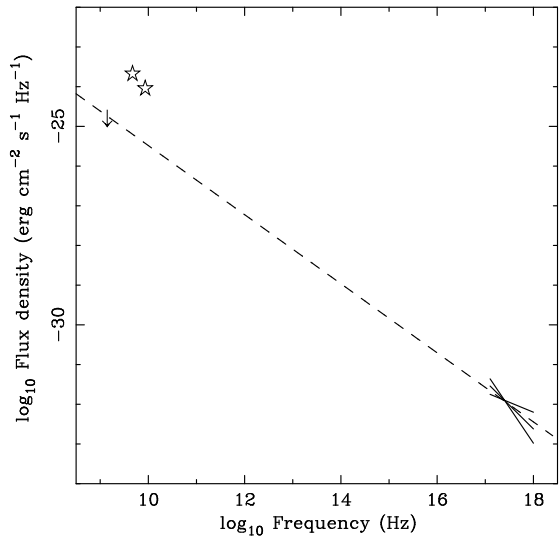}}

\caption{Spectral energy distribution of the northern hot spot showing the 1.4 GHz radio flux of the compact source (upper limit), the 4.7 GHz and 8.6 GHz fluxes of the more extended emission (stars) and the X-ray spectrum (``bow-tie'').  The dashed line shows a power law $S_\nu \propto \nu^{-0.87}$ \citep[the more diffuse radio emission associated with the northern hot spot was estimated to have a spectral slope of 0.87 by][]{1998MNRAS.296..701F}.}

\label{fig:nhs}
\end{figure}

The spectral energy distribution of the northern hot spot is shown in
Fig.~\ref{fig:nhs}.  The northern lobe hot spot was not detected in our VLBI
observation with an estimated upper limit to the total flux density of the
compact emission of approximately 25 mJy (see section \ref{sec:morph:nhs}). 
The more diffuse 4.7~GHz radio emission associated with the northern hot spot
has a flux density of 210~mJy and a spectral slope of $\alpha = 0.87$
\citep[the spectral indices measured from the lower resolution radio maps should be interpreted with caution because the observations were not made with matched arrays;][]{1998MNRAS.296..701F}.  A power law model provides a good description of the spectrum, although if the emitting region is spatially unresolved its 1.4 GHz radio flux density should be just below the upper limit.

\begin{figure}
\centerline{\includegraphics[scale=0.4,angle=270]{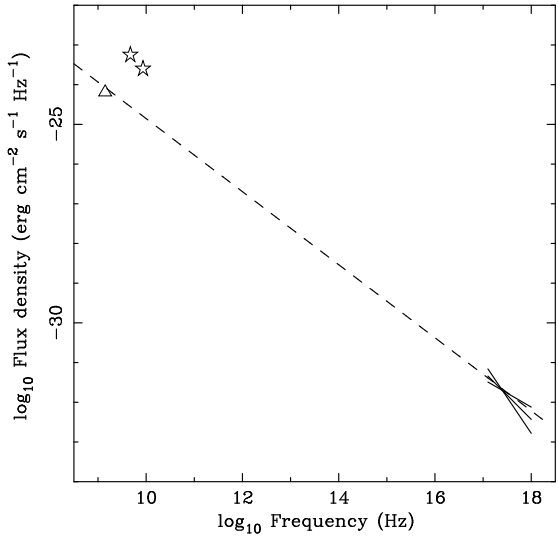}}

\caption{Spectral energy distribution of the southern hot spot showing the 1.4 GHz radio flux of the brightest compact knot (triangle), the 4.7 GHz and 8.6 GHz fluxes of the more extended emission (stars) and the X-ray spectrum (``bow-tie'').  The dashed line shows a power law $S_\nu \propto \nu^{-0.92}$ \citep[the more diffuse radio emission associated with the southern hot spot was estimated to have a spectral slope of 0.92 by][]{1998MNRAS.296..701F}.}

\label{fig:shs}
\end{figure}

The spectral energy distribution of the southern hot spot is shown in
Fig.~\ref{fig:shs}.  The 1.4~GHz flux density of the brightest knot in our VLBI
image of the southern hot spot is 65~mJy, which is significantly lower than the
4.7~GHz flux density of 560~mJy reported by  \citet{1998MNRAS.296..701F}.  This
difference in flux densities is just due to the significant difference in beam
sizes.  If the X-ray emission is associated with the compact knots in the VLBI
image we find that a power law model provides an excellent description of the
spectrum, with a spectral slope of $\alpha = 0.92$ which is equal to the radio
spectral slope estimated by \citet{1998MNRAS.296..701F}.  The 1.4 GHz radio to X-ray spectral slope is $\alpha_{\rm rx} = 0.94$.

\subsection{The jet/cloud interaction} \label{sec:disc:jci}

As shown in Figs. \ref{fig:cxo_nuc_hic} and \ref{fig:cxo_hst_vlbi}, the jet/cloud interaction region in PKS 2153--69 has a complex multi-wavelength
structure.  The eastern component of the region (Figs. \ref{fig:cxo_nuc_hic} and \ref{fig:cxo_hst_vlbi}) consists of an ensemble of highly ionized filamentary clouds, with the highest ionization clouds closest to the nucleus \citep{1988MNRAS.235..403T}.  Spectra of these clouds  also reveal the presence of a blue polarized continuum that may be scattered light from the active nucleus \citep{2001MNRAS.328..848V, 1998MNRAS.296..701F}.  In Section \ref{sec:hic_spec} we concluded that the X-rays observed from the eastern component may come from a thermal plasma with $kT = 0.22 \keV$, plus some additional weaker high temperature or non-thermal emission.  An alternative, discussed in Section \ref{sec:dust}, is that the X-rays, like the optical continuum, are light from the ``blazar'' nucleus which has been scattered by dust into our line of sight.  The VLBI jet (Section \ref{sec:morph_nuc}) is seen to be highly aligned with the set of clouds that make up the eastern component \citep{2002ApJS..141..311T, 1996AJ....111..718T}.

The western component of the interaction region (Fig. \ref{fig:cxo_hst_vlbi})
is dominated by a feature that has a red optical  continuum, in contrast to the
blue continuum of the eastern component.  Radio emission has been observed
coincident with the western component by \cite{1998MNRAS.296..701F} and here we
have reported the detection of X-ray emission (Section \ref{sec:hic_spec}),
although it is unclear if this emission is thermal or non-thermal in nature. 
The position angle connecting the nucleus and the western component is
approximately $10\degmark$ less than the position angle of the VLBI jet
direction that intersects the center of the eastern component
\citep{1998MNRAS.296..701F}.

Confronted with these facts, can we plausibly identify the region within the
ensemble of clouds where the primary jet interaction occurs?

For a start, it appears unlikely that the radio jet has an opening angle that
encompasses the entire cloud complex, both the eastern and western components. 
The opening angle of the nuclear jet, from VLBI Space Observatory Program
observations, is very small, less than a few degrees
\citep{2002ApJS..141..311T}.  Also, the diameter of the jet at the southern
lobe hot spot is presumably less than the measured diameter of the hot spot itself,
approximately 200 pc.  Therefore, it is likely that the primary site of
interaction will be confined to only a small part of the cloud complex.  It
is likely that many shocks exist within the cloud complex as a result of the
jet interaction \citep{1998ApJ...495..680B} and it may be these secondary
shocks that are powering the thermal X-ray emission throughout the region.

\begin{figure}
\centerline{\includegraphics[scale=0.4,angle=270]{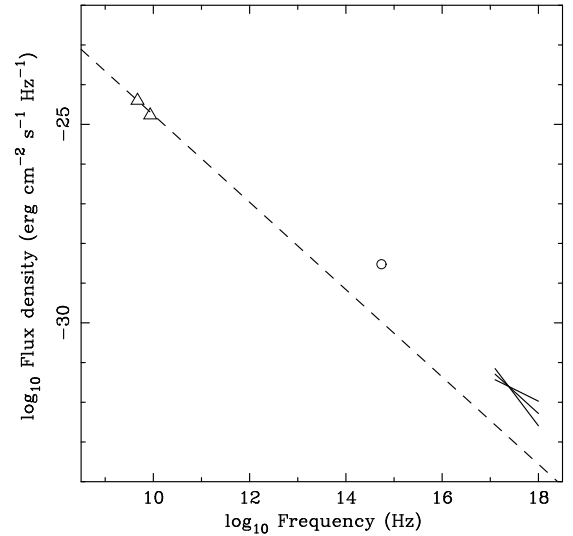}}

\caption{Spectral energy distribution of the western component of the jet/cloud interaction region from radio (triangles) through optical (circle) to
X-ray (``bow-tie'') wavelengths.  The dashed line shows a power law spectrum
$S_\nu \propto \nu^{-\alpha}$ with $\alpha = 1.1$ corresponding to the
estimated spectral slope of the radio emission \citep{1998MNRAS.296..701F}. 
The radio flux densities are taken from \citet{1998MNRAS.296..701F} and the
western component was not detected in our VLBI observation.  The optical flux
density is from \citet{1988Natur.334..591D}. \label{fig:rso}}

\end{figure}

Previously, it has been suggested that the red continuum of the western
component is radiation from a foreground star \citep{1988Natur.334..591D}.  However, the
existence of both radio and X-ray emission coincident with this component
almost certainly marks it as integral to the cloud complex. 
\citet{2001MNRAS.328..848V} give an alternative explanation for the red
continuum: they suggest that it results from large dust grains scattering red
light from the active nucleus, the small dust grains in this component having
been destroyed by shocks in a jet/cloud interaction.  According to this
explanation, the blue continuum of the eastern component is a result of scattering of blue light from the nucleus by small dust grains that are
preserved because the eastern component is not as heavily shocked as the
western component.  Identification of the western component as the primary jet
interaction site, via this argument, is consistent with the presence of radio
synchrotron emission (increased synchrotron emissivity due to enhanced electron
energy and compressed magnetic field in the shocked region) and X-ray
emission.  Fig. \ref{fig:rso} shows the radio to X-ray spectrum for the western
component.  This spectrum could be plausibly fitted with a power law with the spectral slope given by \citet{1998MNRAS.296..701F}, and an additional
optical component of red light from the nucleus scattered by large dust
grains \citep{2001MNRAS.328..848V}.  There is also thermal X-ray emission
from a hot plasma (see Section \ref{sec:hic_spec}).  The
scattered optical and thermal X-ray emission dominate, and are at least an
order of magnitude stronger than the synchrotron emission extrapolated to optical and X-ray wavelengths.

\begin{figure}
\centerline{\includegraphics[scale=0.4,angle=270]{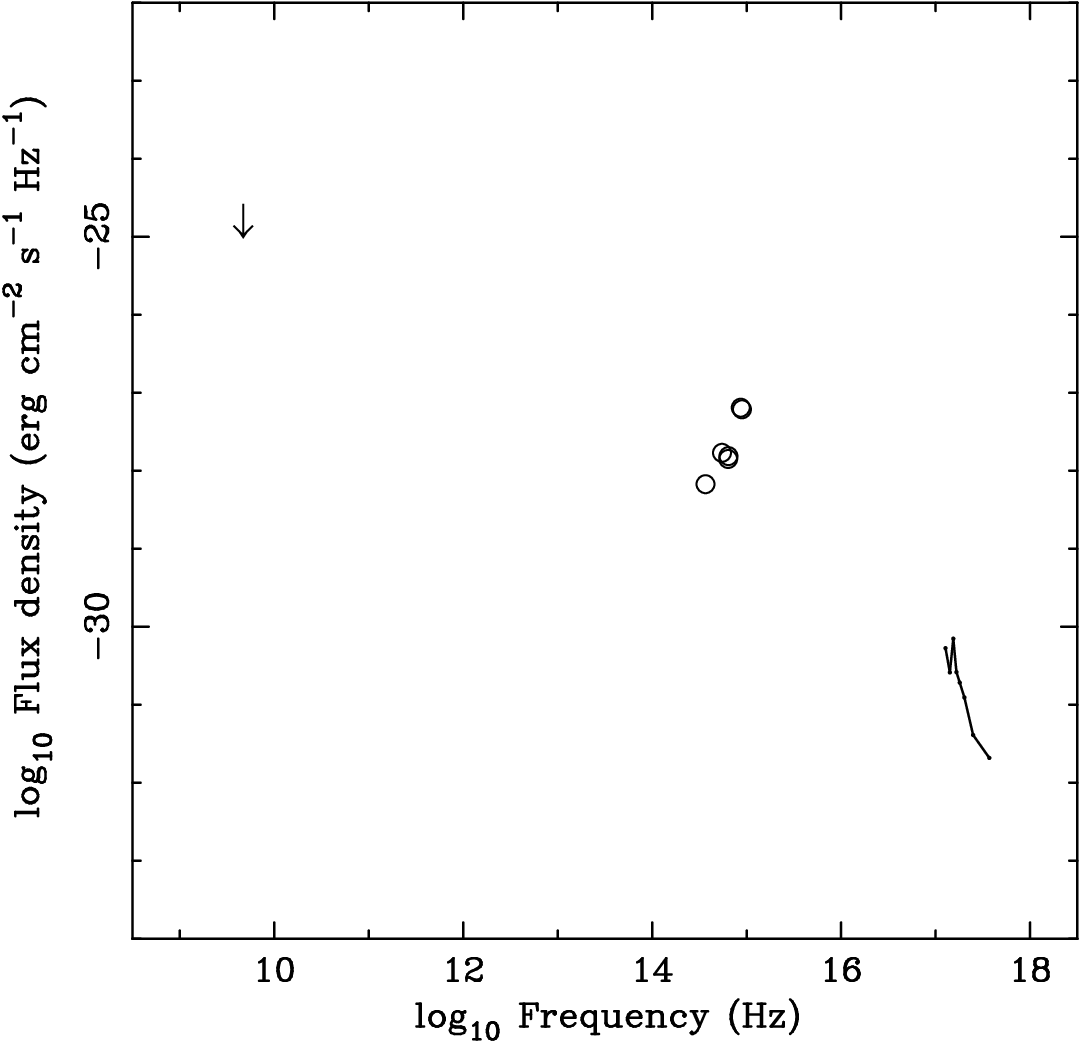}}

\caption{Spectral energy distribution of the eastern component of the jet/cloud interaction region from radio (upper limit) through optical (circles) to
X-ray (solid line) wavelengths.  The radio upper limit was estimated from
\citet{1998MNRAS.296..701F}, and the optical continuum flux densities (which
follow a $\nu^{+3}$ law) are from \citet{1988Natur.334..591D}.  The X-ray
spectrum shows the data unfolded using the best fit model given in Table
\ref{tab:spec}.  The optical continuum emission is $\simeq 12\pct$ linearly
polarized \citep{1988Natur.334..591D} and is probably radiation that has been
beamed from the nucleus and scattered into our line of sight by dust in the cloud.  The X-ray emission may be similarly dust scattered nuclear light or thermal plasma emission. \label{fig:hic}}

\end{figure}

However, there is a problem with identifying the western component as the primary jet interaction site.  All VLBI imaging observations of the nuclear radio jet show that it is highly aligned with the eastern component, not the western component.  Further, the strongest X-ray emission from the cloud complex comes from the eastern component.  The radio to X-ray spectrum of the eastern component is shown in Fig. \ref{fig:hic}.  A solution to this problem might be that radiation beamed from the nucleus along the VLBI jet direction is being scattered by dust into our line of sight giving rise to the observed optical and X-ray emission.  We consider this possibility below.

\subsection{Dust Scattering} \label{sec:dust}

The eastern component of the jet/cloud interaction region is illuminated by radiation beamed along the VLBI jet direction from the nucleus, and some of this radiation may be scattered by dust into our line of sight.  The polarized, blue optical continuum is thought to be the result of dust scattering \citep{2001MNRAS.328..848V}, and it is possible that the X-ray emission is also the result of dust scattering.  If the radiation beamed from the nucleus has a BL Lac spectrum the ratio of the optical to X-ray flux densities seen by the cloud will be approximately $L_{\nu, {\rm opt}} / L_{\nu, {\rm x}} = 2 \times 10^2$ to $2 \times 10^3$ \citep{1998MNRAS.299..433F}.  The differential scattering cross section for blue light is $(d\sigma / d\Omega)_{\rm opt} \simeq 2 \times 10^{-22} \cmsq \psr {\rm H}^{-1}$ \citep{2003ApJ...598.1017D} for $\theta_{\rm s} \approxlt 20\degmark$, where ${\rm H}^{_-1}$ denotes per Hydrogen nucleon.  Here we have averaged the results for the Milky Way and the LMC plotted in Fig.~3 of \citet{2003ApJ...598.1017D}.  The differential scattering cross section for 0.45 keV X-rays is $(d\sigma / d\Omega)_{0.45 \keV} \simeq 1.5 \times 10^{-20} (\theta_{\rm s} / 1\degmark)^{-3} \cmsq \psr {\rm H}^{-1}$ for $\theta_{\rm s} \approxgt 0.3\degmark$, where $\theta_{\rm s}$ is the scattering angle and the $-3$ exponent has been estimated from Fig.~8 of \citet{2003ApJ...594..347D}.  These scattering cross sections assume a Milky Way distribution of dust grain sizes and composition.  The observed ratio of the optical to X-ray flux density of the eastern component is approximately $F_{\nu, {\rm opt}}/F_{\nu, {\rm x}} = 10^3$, and this constrains the scattering angle since
\begin{equation}
\frac{F_{\nu, {\rm opt}}}{F_{\nu, {\rm x}}} \simeq \frac{L_{\nu, {\rm opt}}}{L_{\nu, {\rm x}}} \frac{(d\sigma / d\Omega)_{\rm opt}}{(d\sigma / d\Omega)_{\rm x}} \simeq 1000 \left(\frac{L_{\nu, {\rm opt}}/L_{\nu, {\rm x}}}{2 \times 10^2}\right) \left(\frac{\theta_{\rm s}}{6\degmark}\right)^3,
\end{equation}
giving $\theta_{\rm s} = 3\degmark - 6\degmark$ for $L_{\nu, {\rm opt}} / L_{\nu, {\rm x}} = 2 \times 10^2 - 2 \times 10^3$, respectively.  Given the uncertainties in the spectral energy distribution of the radiation beamed from the nucleus, the distribution of dust grain sizes and the scattering angle it may be that the X-ray emission from the eastern component represents dust scattering of nuclear light.  However, the predicted X-ray intensity is only comparable to that observed if $\theta_{\rm s}$ is only a few degrees and there is no evidence that the radio axis of PKS 2153--69 is at such a small angle to the line of sight.  If the X-rays \emph{are} dust scattered, this cloud in PKS 2153--69 would be the first example of this process in an extragalactic context.

\subsection{Electron Scattering}

Radiation beamed from the nucleus towards the eastern component of the jet/cloud interaction region may be electron scattered into our line of sight as long as the scattering medium is highly ionized, so that soft X-ray absorption is negligible, and has a sufficiently large column density.  We consider electron scattering to be unlikely, however, because the spectrum of the eastern region of the jet/cloud interaction region ($\Gamma > 4$) is much softer and clearly not the same as the spectrum of the nucleus ($\Gamma = 1.75$).  The cloud complex might see a much softer spectrum than we do, which is possible if soft X-rays from the nucleus are heavily obscured in our direction but propagate freely to the scattering cloud.  The X-ray spectrum of the nucleus, however, does not show additional absorption in excess of the Galactic column, so it is unlikely that there is such an obscured soft X-ray source in the nucleus.

\subsection{Jet Precession} \label{sec:disc:precession}

A solution to the problem of having two different sites of interaction within the cloud (one for the radiation beam and one for the particle beam) is jet precession.  If the jet is slow ($v_{\rm j} \ll c$)
and the jet precesses across the cloud complex, a time delay between the photon
beam from the nucleus and the particle beam reaching a given part of the cloud
complex will exist.  Such a scenario may explain the fact that the western
component has radio, optical, and X-ray emission consistent with a jet/cloud
interaction region, while the eastern component has a high degree of ionization
(photoionization from the beamed nuclear radiation, plus a possible contribution from
any secondary shocks in the cloud due to the interaction in the western
component), and scattered nuclear continuum light.

For this hypothesis to be viable, the jet precession rate must be slow enough that, at present, the direction of the nuclear jet (which is presumed to align with the nuclear photon beam and hence with the photoionized eastern component of the cloud complex) is only 15 -- 20$\degmark$ in projection from the direction between the nucleus and the western component of the cloud complex, where the jet's interaction with the cloud is presumed to take place.  The light travel time between the nucleus and the cloud complex is approximately $20,000 / \sin \theta$ years ($\theta$ is the angle between the jet and our line of sight).  During this time the nuclear jet can only have changed its true orientation by less than a few degrees (the angular extent of the eastern cloud as seen from the nucleus), in order to remain in near alignment with the eastern component, giving a rate of change for the jet orientation of $< 2 \times 10^{-4} \sin \theta$ $\degmark$ yr$^{-1}$.  Thus, at this rate of change, the jet material that is currently impacting the western component of the cloud system was ejected from the nucleus at least $40,000 / \sin \theta$ years ago.  The inferred speed of the jet is therefore $< 0.5$c (averaged over the journey from the nucleus to the cloud system).  The jet speed inferred from equation \ref{eqn:j_cj} is higher than this for some values of $\theta$, but at its origin in the nucleus the jet may well be faster than it is several kpc away at the clouds, due to  deceleration.  

Also the sense of the jet precession (from western to eastern component), if
mirrored on the opposite side of the nucleus, follows a curved feature evident
in the radio emission that ends at the current terminus of the southern jet in
an active lobe hot spot (as seen from the VLBI images).  Alternatively this
curved  feature in the southern lobe could be interpreted as backflow from the
lobe hot spot.  The jet precession hypothesis predicts that the current nuclear
counter-jet (not visible due to beaming effects --- see equation \ref{eqn:j_cj})
should not align with the southern lobe hot spot but should be pointed somewhat
to the west of the hot spot.  If we assume that the nuclear counter-jet is
diametrically opposed to the observed nuclear jet (which is likely given that
this is the case for every counter-jet yet observed) then the ATCA images of
\cite{1998MNRAS.296..701F} agree with the prediction.  The counter-jet position
angle would then be approximately $19\degmark$ greater than the position angle joining the nucleus and the southern lobe hot spot \citep{1998MNRAS.296..701F}.  At a jet speed of 0.5c, this 19$^{\circ}$ offset represents a jet orientation rate of change very close to that estimated for the northern jet at the cloud complex, and in the same sense.  The P.A. of the northern hot spot with respect to the nucleus is lower than the P.A. of the jet/cloud interaction site (the western component) giving an overall ``S-shape'' to the radio source.

\begin{figure}
\centerline{\includegraphics[scale=0.55,angle=270]{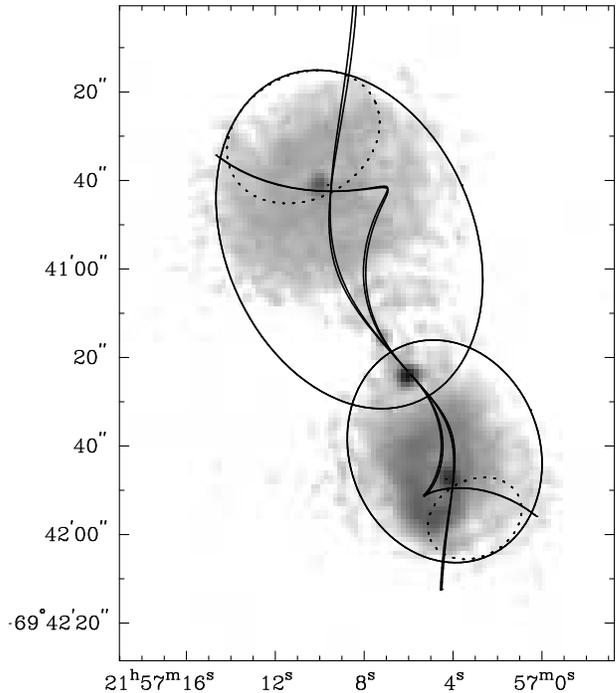}}

\caption{A jet precession model for PKS 2153--69 overlaid on a 4.7~GHz radio
map \citep{1998MNRAS.296..701F}.  The radio lobes are approximated by spheroids
(solid ellipses) and the lobe hot spots are assumed to lie on the surface of
the spheroids at the intersection of the spheroid surfaces and the cone formed
by the precessing jet direction (dotted ellipses).  Four possible solutions are
shown by the curved lines (since there are two possible precession directions
and two possible azimuthal positions of the pc-scale jet that give the same
projected position).  The model and its parameters are described in Section
\ref{sec:disc:precession} \label{fig:precession}}

\end{figure}

We can extend this argument to the position of the northern hot spot, taking into account the finite jet speed, light travel time and projection effects.  Our model consists of symmetric back-to-back ballistic jets moving at constant speed $v_{\rm j}$ and precessing at a rate $\Omega$ with the precessing jet direction forming a cone of half-opening angle $\theta_{\rm c}$.  The symmetry axis of the cone is inclined at an angle $i$ to our line of sight.  The radio lobes are approximated by spheroids and the hot spots are assumed to lie on the surfaces of the spheroids.  An additional parameter is the present day azimuthal phase of the jet $\phi$.  Given an inclination $i$ and a cone half-opening angle $\theta_{\rm c}$ we can solve for $\Omega$, $v_{\rm j}$ and $\phi$ from trigonometric and light travel time considerations.  For all possible values of $i$ and $\theta_{\rm c}$ we compare the predicted jet path with the observed hot spot positions and pick solutions that have the smallest deviation.  There is not a unique solution for $i$ and $\theta_{\rm c}$ so we choose the one with the largest $v_{\rm j}$ given the VLBI beaming constraints.  These precession solutions for both possible values of $\phi$ and both positive and negative $\Omega$, are shown in Fig.~\ref{fig:precession}.  In this model $i = 40.5\degmark$, $\theta_{\rm c} = 11.6\degmark$, $\beta_{\rm j} = v_{\rm j} / c \simeq 0.3$ and the current inclination angle of the pc-scale jet to our line of sight is $\simeq 40\degmark$.  Two solutions have precession rates of $\Omega \simeq -2 \times 10^{-4}$ $\degmark$ yr$^{-1}$ and two have $\Omega \simeq 5 \times 10^{-4}$ $\degmark$ yr$^{-1}$, where positive $\Omega$ implies precession in a clockwise sense.  The slower precession rates give the ``straighter'' curves in Fig.~\ref{fig:precession}, and are in good agreement with the precession rates expected from the jet orientation rates of change estimated above and the geometry adopted for the precession model.

\subsection{Jet Deflection}

The idea of jet precession, as outlined in the previous section, seems to plausibly explain a great many of the observed features of PKS 2153--69 at X-ray, radio, and optical wavelengths.  Previously, \citet{1997A&A...327..550T} suggested that the radio morphology of PKS 2153--69 is affected by deflection of the northern jet at the site of the jet/cloud interaction, causing the observed difference between the nuclear jet position angle and the position angle joining the nucleus and the northern lobe hot spot.  For the reasons stated above, we feel that jet precession gives a better overall explanation of the multi-wavelength source structure, but we briefly revisit the alternative jet deflection explanation, using the previously unavailable constraints from the X-ray observations.

Under this interpretation, the jet from the nucleus must be deflected through $\sim 20\degmark$ (projected onto the plane of the sky) by the cloud before continuing on to the northern hot spot \citep{1997A&A...327..550T, 2002ApJS..141..311T}.  Using some of our measured quantities it is possible to make an estimate of the jet speed at the cloud complex, following the calculations of \cite{1998ApJ...495..680B} and \cite{2003NewAR..47..569K}.  Using the expressions for the jet power in \cite{1998ApJ...495..680B} we estimate a jet power, $F_{\rm j} = 5 \times 10^{41} - 5 \times 10^{43} \ergps$ (assuming $\kappa_\nu$, the ratio of monochromatic power at frequency $\nu$ to total jet flux, lies in the range $10^{-11} < \kappa_{\nu} < 10^{-9}$, and the radio flux density at 1.4 GHz is half of the total flux density of the entire source at 1.4 GHz, to reflect the power in only the northern jet).  This compares to the $\sim 2 \times 10^{42}$ erg/s estimated in Section \ref{sec:dis_halo}.   The jet speed can then be estimated as,

\begin{equation}
v_{\rm j} \sim \frac{0.5 f(M_{\rm j},\Delta \theta) F_{\rm j}
v_s}{L_X},
\end{equation}

where $L_X$ is the X-ray luminosity associated with the cloud complex ($L_X =
1.1 \times 10^{41} \ergps$ between 0.5 and 7 keV), $v_{\rm s}$ is the shock velocity (which we will take to be the shock velocity of $430 \kmps$ [equation \ref{eq:shock}] in the eastern part of the cloud complex assuming a thermal interpretation of its X-ray emission) and $f(M_{\rm j},\Delta \theta)$ is  described and calculated in \cite{1998ApJ...495..680B} for a non-relativistic jet, depending on the jet Mach number, $M_{\rm j}$, and the jet deflection angle, $\Delta \theta$.  For PKS 2153$-$69, taking $\Delta \theta = 20\degmark$ \citep{1996AJ....111..718T} and $f(M_{\rm j},\Delta \theta) \sim 0.5$, we estimate that $v_{\rm j} < 0.2c$.

It is interesting to compare these mildly relativistic jet speeds with the
estimate made by \cite{1996AJ....111..718T} starting from the assumption of a
{\it relativistic} jet at the clouds, resulting in $v_{\rm j}>0.9c$.  The
calculation of \cite{1996AJ....111..718T} does not use the X-ray emission as a
constraint on the jet energetics (because no suitable observations were 
available at that point) and considers simple oblique shocks in a relativistic
jet.

Other interpretations of S-shaped radio sources include jet bending by the thermal pressure of a hot, gaseous atmosphere \citep{1981ApJ...249...40H} or by the ram pressure of rotating interstellar gas \citep{1982ApJ...263..576W}.

\section{Conclusions}

In this paper we present the results of Chandra and 1.4 GHz LBA observations of
the radio galaxy PKS 2153--69 and its environment.  Our conclusions are
summarized below.

1.  The Chandra image reveals a roughly spherical halo of hot gas extending out
to $30\kpc$ around PKS 2153--69 with a temperature of $kT = 1.01 \keV$, a 0.5
-- 2 keV luminosity of $7 \times 10^{41} \ergps$ and a total gas mass of $3
\times 10^{10} \Msun$ (the total gravitating mass is likely much larger).  Two
depressions in the surface brightness of the X-ray halo correspond to the large
scale radio lobes, and assuming these are cavities inflated by the jet we infer
a total jet power (both sides) of $4 \times 10^{42} \ergps$.

2.  Both the northern and southern lobe hot spots are detected by Chandra.  In
addition, the southern hot spot is detected in the 1.4~GHz VLBI observation
providing the highest linear resolution image of a radio lobe hot spot to date. 
The northern hot spot is not detected in the VLBI observation.

3.  The X-ray spectrum of the nucleus is well described by a power law of
photon index $\Gamma = 1.75$ absorbed by the Galactic column, with a 2 -- 10
keV X-ray luminosity of $9 \times 10^{42} \ergps$.  This photon index is typical of a type 1 active galaxy.  The 1.4 GHz VLBI image of the nucleus shows a jet on $0\farcs1$ scales along a P.A. of $45\degmark$.  The absence of a detectable counter-jet constrains the jet speed $v_{\rm j}$ ($\beta_{\rm j} = v_{\rm j}/c$) and inclination angle $\theta$ to be $\beta_{\rm j} \cos \theta > 0.37$ on pc scales.

4.  Approximately $10\arcsec$ northeast of the nucleus, at a position angle
intermediate between that of the northern hot spot and the direction of the
nuclear jet, X-ray emission is detected from an extra-nuclear cloud. The X-ray
emission from the cloud can be divided into two regions, an unresolved western
component associated with a knot of radio emission, and a spatially
resolved eastern component associated with highly ionized optical line-emitting
clouds.  The radio spectrum of the western component is consistent
with synchrotron emission, with the optical continuum
being predominantly red light scattered from nuclear light impinging on the cloud by large dust grains
and the X-ray emission being from a hot thermal plasma.  The X-ray emission of
the eastern component has a 0.5 -- 2 keV luminosity of $7 \times 10^{40}
\ergps$ and is very soft ($\Gamma > 4$ or $kT \simeq 0.2 \keV$).  The eastern component is probably being photoionized by radiation beamed from the nucleus with the polarized blue optical continuum and soft X-ray emission being beamed radiation from the nucleus scattered into our line of sight by dust in the clouds.  The interpretation of this soft X-ray emission as dust-scattered nuclear light is speculative and appears to require that the radio axis of PKS 2153--69 is within several degrees of our line of sight.  If correct, this is the first detection of dust-scattered X-rays in an extragalactic object.  The 1.4 GHz VLBI observation did not detect compact radio emission from the extra-nuclear cloud.

5.  We account for the progressive increase of P.A. from the northern hot spot to the
western region of the extra-nuclear cloud to the pc-scale radio jet in terms of
a jet precession model in which the particle beam interacts with the western
component of the jet/cloud interaction region and the radiation beam
interacts with the eastern component.  The jet speed is estimated to be $<0.5 c$ and the precession rate to be $\simeq 2 \times 10^{-4}$ $\degmark$
yr$^{-1}$.  The model is consistent with the location of the southern hot spot, which gives the source an overall ``S-shape''.  We discuss the alternative possibility of jet deflection.

6.  X-ray emission from the nearby radio galaxy MRC 2153-699 is detected in the Chandra image.

Our findings are consistent with and complimentary to the presentation of the X-ray observations in a preprint by \citet{2004astro.ph..9272L} that appeared during the refereeing of our paper.

\section{Acknowledgments}

We thank R. Fosbury for providing electronic copies of his radio maps.  This work was supported by NASA through contract NAS8-01129 and grants NAG8-1027 and NAG5-13065, the Chandra Fellowship Award Number PF3-40026 issued by the Chandra X-ray Observatory Center which is operated by the Smithsonian Astrophysical Observatory for and on behalf of the NASA under contract NAS8-39073, and by a grant from the Research and Development Grants Scheme of the Swinburne University of Technology.  The Australia Telescope is funded by the Australian Commonwealth Government for operation as a national facility  managed by the CSIRO.

\bibliographystyle{apj}
\bibliography{apj-jour,bibliography}

\begin{thebibliography}{40}
\expandafter\ifx\csname natexlab\endcsname\relax\def\natexlab#1{#1}\fi

\bibitem[{{Arnaud}(1996)}]{1996adass...5...17A}
{Arnaud}, K.~A. 1996, in ASP Conf. Ser. 101: Astronomical Data Analysis
  Software and Systems V, Vol.~5, 17

\bibitem[{{Bennett} {et~al.}(2003){Bennett}, {Halpern}, {Hinshaw}, {Jarosik},
  {Kogut}, {Limon}, {Meyer}, {Page}, {Spergel}, {Tucker}, {Wollack}, {Wright},
  {Barnes}, {Greason}, {Hill}, {Komatsu}, {Nolta}, {Odegard}, {Peiris},
  {Verde}, \& {Weiland}}]{2003ApJS..148....1B}
{Bennett}, C.~L., {Halpern}, M., {Hinshaw}, G., {Jarosik}, N., {Kogut}, A.,
  {Limon}, M., {Meyer}, S.~S., {Page}, L., {Spergel}, D.~N., {Tucker}, G.~S.,
  {Wollack}, E., {Wright}, E.~L., {Barnes}, C., {Greason}, M.~R., {Hill},
  R.~S., {Komatsu}, E., {Nolta}, M.~R., {Odegard}, N., {Peiris}, H.~V.,
  {Verde}, L., \& {Weiland}, J.~L. 2003, \apjs, 148, 1

\bibitem[{{Bicknell}(1994)}]{1994ApJ...422..542B}
{Bicknell}, G.~V. 1994, \apj, 422, 542

\bibitem[{{Bicknell} {et~al.}(1998){Bicknell}, {Dopita}, {Tsvetanov}, \&
  {Sutherland}}]{1998ApJ...495..680B}
{Bicknell}, G.~V., {Dopita}, M.~A., {Tsvetanov}, Z.~I., \& {Sutherland}, R.~S.
  1998, \apj, 495, 680

\bibitem[{{Carilli} {et~al.}(1991){Carilli}, {Perley}, {Dreher}, \&
  {Leahy}}]{1991ApJ...383..554C}
{Carilli}, C.~L., {Perley}, R.~A., {Dreher}, J.~W., \& {Leahy}, J.~P. 1991,
  \apj, 383, 554

\bibitem[{{Davis}(2001)}]{2001ApJ...562..575D}
{Davis}, J.~E. 2001, \apj, 562, 575

\bibitem[{{De Young}(2004)}]{de_young_04}
{De Young}, D.~S. 2004, to appear in ``X-Ray and Radio Connections'', eds. L.
  Sjouwerman and K. Dyer

\bibitem[{{di Serego Alighieri} {et~al.}(1988){di Serego Alighieri},
  {Courvoisier}, {Fosbury}, {Tadhunter}, \& {Binette}}]{1988Natur.334..591D}
{di Serego Alighieri}, S., {Courvoisier}, T.~J.-L., {Fosbury}, R.~A.~E.,
  {Tadhunter}, C.~N., \& {Binette}, L. 1988, \nat, 334, 591

\bibitem[{{Dickey} \& {Lockman}(1990)}]{1990ARA&A..28..215D}
{Dickey}, J.~M. \& {Lockman}, F.~J. 1990, \araa, 28, 215

\bibitem[{{Draine}(2003)}]{2003ApJ...598.1017D}
{Draine}, B.~T. 2003, \apj, 598, 1017

\bibitem[{{Draine} \& {Tan}(2003)}]{2003ApJ...594..347D}
{Draine}, B.~T. \& {Tan}, J.~C. 2003, \apj, 594, 347

\bibitem[{{Fosbury} {et~al.}(1990){Fosbury}, {Di Serego Alighieri},
  {Courvoisier}, {Snijders}, {Tadhunter}, {Walsh}, \&
  {Wilson}}]{1990iuea.rept..513F}
{Fosbury}, R.~A.~E., {Di Serego Alighieri}, S., {Courvoisier}, T.~J.-L.,
  {Snijders}, M.~A.~J., {Tadhunter}, C.~N., {Walsh}, J., \& {Wilson}, W. 1990,
  in Evolution in Astrophysics: IUE Astronomy in the Era of New Space Missions,
  513--516

\bibitem[{{Fosbury} {et~al.}(1998){Fosbury}, {Morganti}, {Wilson}, {Ekers}, {di
  Serego Alighieri}, \& {Tadhunter}}]{1998MNRAS.296..701F}
{Fosbury}, R.~A.~E., {Morganti}, R., {Wilson}, W., {Ekers}, R.~D., {di Serego
  Alighieri}, S., \& {Tadhunter}, C.~N. 1998, \mnras, 296, 701

\bibitem[{{Fossati} {et~al.}(1998){Fossati}, {Maraschi}, {Celotti}, {Comastri},
  \& {Ghisellini}}]{1998MNRAS.299..433F}
{Fossati}, G., {Maraschi}, L., {Celotti}, A., {Comastri}, A., \& {Ghisellini},
  G. 1998, \mnras, 299, 433

\bibitem[{{Hardcastle} {et~al.}(2003){Hardcastle}, {Worrall}, {Kraft},
  {Forman}, {Jones}, \& {Murray}}]{2003ApJ...593..169H}
{Hardcastle}, M.~J., {Worrall}, D.~M., {Kraft}, R.~P., {Forman}, W.~R.,
  {Jones}, C., \& {Murray}, S.~S. 2003, \apj, 593, 169

\bibitem[{{Henriksen} {et~al.}(1981){Henriksen}, {Vallee}, \&
  {Bridle}}]{1981ApJ...249...40H}
{Henriksen}, R.~N., {Vallee}, J.~P., \& {Bridle}, A.~H. 1981, \apj, 249, 40

\bibitem[{{Hughes} {et~al.}(2002){Hughes}, {Miller}, \&
  {Duncan}}]{2002ApJ...572..713H}
{Hughes}, P.~A., {Miller}, M.~A., \& {Duncan}, G.~C. 2002, \apj, 572, 713

\bibitem[{{Jones} \& {McAdam}(1992)}]{1992ApJS...80..137J}
{Jones}, P.~A. \& {McAdam}, W.~B. 1992, \apjs, 80, 137

\bibitem[{{Kaastra}(1992)}]{1992kaastra}
{Kaastra}, J.~S. 1992, in Internal SRON-Leiden Report, updated version 2.0

\bibitem[{{Kadler} {et~al.}(2003){Kadler}, {Ros}, {Kerp}, {Falcke}, {Zensus},
  {Pogge}, \& {Bicknell}}]{2003NewAR..47..569K}
{Kadler}, M., {Ros}, E., {Kerp}, J., {Falcke}, H., {Zensus}, J.~A., {Pogge},
  R.~W., \& {Bicknell}, G.~V. 2003, New Astronomy Review, 47, 569

\bibitem[{{Landau} \& {Lifshitz}(1959)}]{1959flme.book.....L}
{Landau}, L.~D. \& {Lifshitz}, E.~M. 1959, {Fluid mechanics} (Course of
  theoretical physics, Oxford: Pergamon Press, 1959)

\bibitem[{{Liedahl} {et~al.}(1995){Liedahl}, {Osterheld}, \&
  {Goldstein}}]{1995ApJ...438L.115L}
{Liedahl}, D.~A., {Osterheld}, A.~L., \& {Goldstein}, W.~H. 1995, \apjl, 438,
  L115

\bibitem[{{Ly} {et~al.}(2004){Ly}, {De Young}, \&
  {Bechtold}}]{2004astro.ph..9272L}
{Ly}, C., {De Young}, D., \& {Bechtold}, J. 2004, ArXiv Astrophysics e-prints,
  astro-ph/0409272

\bibitem[{{Mewe} {et~al.}(1985){Mewe}, {Gronenschild}, \& {van den
  Oord}}]{1985A&AS...62..197M}
{Mewe}, R., {Gronenschild}, E.~H.~B.~M., \& {van den Oord}, G.~H.~J. 1985,
  \aaps, 62, 197

\bibitem[{{Mewe} {et~al.}(1986){Mewe}, {Lemen}, \& {van den
  Oord}}]{1986A&AS...65..511M}
{Mewe}, R., {Lemen}, J.~R., \& {van den Oord}, G.~H.~J. 1986, \aaps, 65, 511

\bibitem[{{O'Dea} {et~al.}(2002){O'Dea}, {de Vries}, {Koekemoer}, {Baum}, \&
  {Mack}}]{2002RMxAC..13..196O}
{O'Dea}, C.~P., {de Vries}, W.~H., {Koekemoer}, A.~M., {Baum}, S.~A., \&
  {Mack}, J. 2002, in Revista Mexicana de Astronomia y Astrofisica Conference
  Series, 196--202

\bibitem[{{Perley} {et~al.}(1997){Perley}, {R\"oser}, \&
  {Meisenheimer}}]{1997A&A...328...12P}
{Perley}, R.~A., {R\"oser}, H., \& {Meisenheimer}, K. 1997, \aap, 328, 12

\bibitem[{{Sault} {et~al.}(1995){Sault}, {Teuben}, \&
  {Wright}}]{1995adass...4..433S}
{Sault}, R.~J., {Teuben}, P.~J., \& {Wright}, M.~C.~H. 1995, in ASP Conf. Ser.
  77: Astronomical Data Analysis Software and Systems IV, 433

\bibitem[{{Shepherd}(1997)}]{1997adass...6...77S}
{Shepherd}, M.~C. 1997, in ASP Conf. Ser. 125: Astronomical Data Analysis
  Software and Systems VI, 77

\bibitem[{{Snellen} {et~al.}(2003){Snellen}, {Mack}, {Schilizzi}, \&
  {Tschager}}]{2003PASA...20...38S}
{Snellen}, I.~A.~G., {Mack}, K.-H., {Schilizzi}, R.~T., \& {Tschager}, W. 2003,
  Publications of the Astronomical Society of Australia, 20, 38

\bibitem[{{Tadhunter} {et~al.}(1987){Tadhunter}, {Fosbury}, {Binette},
  {Danziger}, \& {Robinson}}]{1987Natur.325..504T}
{Tadhunter}, C.~N., {Fosbury}, R.~A.~E., {Binette}, L., {Danziger}, I.~J., \&
  {Robinson}, A. 1987, \nat, 325, 504

\bibitem[{{Tadhunter} {et~al.}(1988){Tadhunter}, {Fosbury}, {di Serego
  Alighieri}, {Bland}, {Danziger}, {Goss}, {McAdam}, \&
  {Snijders}}]{1988MNRAS.235..403T}
{Tadhunter}, C.~N., {Fosbury}, R.~A.~E., {di Serego Alighieri}, S., {Bland},
  J., {Danziger}, I.~J., {Goss}, W.~M., {McAdam}, W.~B., \& {Snijders},
  M.~A.~J. 1988, \mnras, 235, 403

\bibitem[{{Tingay}(1997)}]{1997A&A...327..550T}
{Tingay}, S.~J. 1997, \aap, 327, 550

\bibitem[{{Tingay} {et~al.}(1996){Tingay}, {Jauncey}, {Reynolds}, {Tzioumis},
  {Migenes}, {Gough}, {Lovell}, {McCulloch}, {Costa}, {Preston}, \&
  {Harbison}}]{1996AJ....111..718T}
{Tingay}, S.~J., {Jauncey}, D.~L., {Reynolds}, J.~E., {Tzioumis}, A.~K.,
  {Migenes}, V., {Gough}, R., {Lovell}, J.~E.~J., {McCulloch}, P.~M., {Costa},
  M.~E., {Preston}, R.~A., \& {Harbison}, P. 1996, \aj, 111, 718

\bibitem[{{Tingay} {et~al.}(2002){Tingay}, {Reynolds}, {Tzioumis}, {Jauncey},
  {Lovell}, {Dodson}, {Costa}, {McCulloch}, {Edwards}, {Hirabayashi}, {Murphy},
  {Preston}, {Piner}, {Nicolson}, {Quick}, {Kobayashi}, \&
  {Shibata}}]{2002ApJS..141..311T}
{Tingay}, S.~J., {Reynolds}, J.~E., {Tzioumis}, A.~K., {Jauncey}, D.~L.,
  {Lovell}, J.~E.~J., {Dodson}, R., {Costa}, M.~E., {McCulloch}, P.~M.,
  {Edwards}, P.~G., {Hirabayashi}, H., {Murphy}, D.~W., {Preston}, R.~A.,
  {Piner}, B.~G., {Nicolson}, G.~D., {Quick}, J.~F.~H., {Kobayashi}, H., \&
  {Shibata}, K.~M. 2002, \apjs, 141, 311

\bibitem[{{Villar-Mart{\'{\i}}n} {et~al.}(2001){Villar-Mart{\'{\i}}n}, {De
  Young}, {Alonso-Herrero}, {Allen}, \& {Binette}}]{2001MNRAS.328..848V}
{Villar-Mart{\'{\i}}n}, M., {De Young}, D., {Alonso-Herrero}, A., {Allen}, M.,
  \& {Binette}, L. 2001, \mnras, 328, 848

\bibitem[{{Wietfeldt} {et~al.}(1996){Wietfeldt}, {Baer}, {Cannon}, {Feil},
  {Jakovina}, {Leone}, {Newby}, \& {Tan}}]{1996ITIM.45(6).923W}
{Wietfeldt}, R.~D., {Baer}, D., {Cannon}, W.~H., {Feil}, G., {Jakovina}, R.,
  {Leone}, P., {Newby}, P.~S., \& {Tan}, H. 1996, IEEE Transactions on
  Instrumentation and Measurement, 45(6), 923

\bibitem[{{Wilson} \& {Ulvestad}(1982)}]{1982ApJ...263..576W}
{Wilson}, A.~S. \& {Ulvestad}, J.~S. 1982, \apj, 263, 576

\bibitem[{{Wilson} {et~al.}(2001){Wilson}, {Young}, \&
  {Shopbell}}]{2001ApJ...547..740W}
{Wilson}, A.~S., {Young}, A.~J., \& {Shopbell}, P.~L. 2001, \apj, 547, 740

\bibitem[{{Wilson} {et~al.}(1996){Wilson}, {Roberts}, \&
  {Davies}}]{1996.....conf...16W}
{Wilson}, W., {Roberts}, P., \& {Davies}, E. 1996, in Proceedings of the 4th
  APT Workshop, ed. E. King, 16

\end{thebibliography}

\end{document}